# Growth and surface alloying of Fe on Pt(997)

Tae-Yon Lee, Samuel Sarbach, Klaus Kuhnke, and Klaus Kern

Max-Planck-Institut für Festkörperforschung, Heisenbergstrasse 1, D-70569 Stuttgart, Germany

**Abstract**

The growth of ultra-thin layers of Fe on the vicinal Pt(997) surface is studied by thermal energy He atom scattering (TEAS) and Auger Electron Spectroscopy (AES) in the temperature range between 175K and 800K. We find three distinct regimes of qualitatively different growth type: Below 450K the formation of a smooth first monolayer, at and above 600K the onset of bulk alloy formation, and at intermediate temperature 500K - 550K the formation of a surface alloy. Monatomic Fe rows are observed to decorate the substrate steps between 175K and 500K. The importance of the high step density is discussed with respect to the promotion of smooth layer growth and with respect to the alloying process and its kinetics.

# I. INTRODUCTION

The preparation of FePt alloy films is of substantial interest for the study of magnetic materials, both from the fundamental and applied point of view. The magnetic properties of thin FePt alloy films, a high coercivity, high saturation magnetization and a large perpendicular anisotropy are favorable with respect to further storage device miniaturization. Alloys with alternatingly stacked Pt and Fe monolayers (based on a $L1_0$ ordering with a fcc (100) surface orientation or $L1_1$ ordering with a fcc (111) surface) are the subject of intense research with respect to preparation methods and their morphological [1] and magnetic properties [2]. Numerous studies on such systems with comparatively large thickness as well as studies of iron oxides [3, 4] on Pt(111) have been published. There are, however, only few studies [5, 6] which address the Fe ultrathin film growth on Pt(111) surfaces and no studies exist for the growth on surfaces vicinal to Pt(111) and its conditions for surface alloy formation. The growth of ultrathin layers on a vicinal surface exhibits several advantageous properties compared to low index surfaces with respect to growth and alloying behavior. First, true one and two dimensional structures can be formed by step edge decoration, suppressed island nucleation on terraces and forced layer-by-layer growth. Second, alloying can be promoted due to a facilitated intermixing near step edges.

Another aspect makes the system important: It forms a basis for future studies of a bulk-ferromagnetic material on a template surface to create nanometer sized patterns. Such structures have attracted much attention with respect to fundamental questions of magnetism at low dimensions [7, 8]. They are a means to study the origin and size dependence of magnetic properties at the nanometer limit, coupling to the substrate and magnetic isotropy in reduced dimensions. It can be expected that a wealth of structures which can be obtained for FePt alloys allows to learn more on fundamental magnetic properties.

In this paper we study the growth of Fe on Pt(997) in the range of 0 ML – 2 ML in-situ (i.e. during deposition) by thermal energy He scattering. The vicinal Pt(997) surface is employed as a template for Fe deposition. The investigation covers a wide temperature range from 200K to 800K. Within this range the transition from Fe epitaxial growth to surface alloying and further to bulk alloying can be observed. We describe the properties of these phases and the temperature range in which they can be grown.

The paper is organized as follows: The experimental aspects of the study are presented in Sec. II. The measurements by He scattering and Auger electron spectroscopy (AES) are presented in Sec.III. The subsections deal with the low temperature growth (Sec. III A) and the medium and high temperature growth (Sec. III B) separately as the latter involves different aspects of Pt-Fe alloying. The paper is summarized in Sec. IV.

## II. EXPERIMENTAL

The He scattering (TEAS) experiments are carried out in ultra-high vacuum at a base pressure below $2\times10^{-10}$ mbar. The set-up allows to select a wide range of scattering geometries so that He diffraction can be studied both at grazing scattering geometry which is sensitive to the step region and at non-grazing geometry which supplies information on processes on the terraces [9]. Fig.1 presents a schematic drawing of the He beam scattering geometry with regard to the investigated surface. The He nozzle source is cooled by a cold finger and the wavelength of the He atoms is stabilized at $\lambda=1.01$Å for all data shown in this paper. The scattering chamber contains in addition an electron beam bombardment evaporator for Fe evaporation in the He scattering position and a cylindrical mirror Auger spectrometer (Omicron CMA 100) which is employed by rotating the sample from the evaporation position. In addition, low energy electron diffraction (LEED) is available as an analytical tool.

The clean Pt (997) surface is prepared by repeated cycles of 1keV $Ar^+$ sputtering at 750 K surface temperature and annealing at 850 K. In order to avoid a freezing-in of defects which are created at the annealing temperature a slow cooling rate (< 40 K/min) is used after the annealing step. The cleanliness of the surface is checked by AES and the ordering of the substrate staircase structure is monitored by the absolute peak intensities and line shapes in the He diffraction pattern. Fe is evaporated from an electron bombardment evaporator (Omicron single EFM evaporator) loaded with a 99.99% purity Fe-rod. A deposition rate of $R \approx 0.1$ ML/min is employed at a background pressure during evaporation of $3\times10^{-10}$ mbar. The deposition rate is calibrated by the He intensity oscillations observed in non-grazing scattering geometry (see section III).

The geometry of the Pt(997) vicinal surface and the resulting He diffraction pattern has been described in detail elsewhere [10]. Fig.1 represents a sketch of the ideal surface geometry. The surface consists of 20.2 Å wide closed-packed terraces separated by steps with a (111) microfacet of one monolayer height, also called B-type steps. In reality, the surface is well ordered with a distribution of step-step distances which exhibits a maximum near 8 close-packed atomic rows and a standard deviation of 1 row [10].

## III. RESULTS AND DISCUSSION

**A. Epitaxial Fe growth on Pt (997)**

Fig.2 shows the reflected He diffraction intensity in grazing scattering geometry measured in-situ during Fe deposition in the substrate temperature range 175 K ≤ T ≤ 800 K. The average terrace on Pt(997) surface provides space for 8 close-packed adsorbate rows and the scattering in grazing incidence geometry is extremely sensitive to the defect density at the step-edge. Thus the intensity maxima at $\theta_{Fe} \approx 0.13$ ML can be attributed to the formation of complete monatomic rows at each step edge similarly as for many other adsorbates [11]. These maxima are observed for low temperatures (175K) and are still observable up to 500K. The adsorbate is thus mobile enough to decorate the substrate step edges already at 175K. The coverage at the maximum of the first peak deviates, however, from the nominal value of 0.13 ML at the lowest temperatures, which can be explained by the activation threshold for corner rounding as reported earlier for the cases of Co, Cu, and Ag on Pt(997) [12, 13]: Below 250 K, the diffusion barrier between a Fe atom site in the second Fe row at the step edge and a site at the uncovered Pt step edge is higher than the thermal energy.

Fig.3 shows the He diffraction intensity in non-grazing scattering geometry measured during Fe deposition for the substrate temperature range 200 K ≤ T ≤ 700 K. In this geometry information on the morphology and the processes on the terraces can be obtained. We first focus on the range between 200K and 450K: A very prominent intensity maximum appears which indicates the formation of a complete monolayer before adsorption in the second layer begins. We use this maximum at low temperature for a consistent coverage and evaporation rate calibration of our study. The shoulder at 0.2 ML coverage is again a signature of the row formation discussed above. However, in contrast to the data at grazing geometry it can not be interpreted in a straightforward manner: The feature in Fig.3 disappears between T = 300K and T = 450K while Fig.2 demonstrates that the formation of a monatomic row persists all the way up to 500K. At 400K a second intensity maximum near 2ML coverage is observed which suggests that the layer-by-layer growth continues also in the second layer. Maxima for the completion of further layers were not observed, except under conditions when the residual gas pressure was unusually high (probably a surfactant induced behavior which was not studied in detail).

The growth mode established from these results is corroborated by measurements shown in Fig.4 and Fig.5. The diffraction scans in Fig.4 are recorded after depositing the indicated Fe coverages on clean Pt(997) at 350K. Up to a coverage of 0.71 ML the diffraction peaks of a clean Pt(997) surface are preserved although their intensity decreases with increasing coverage. At a coverage of 1.5 ML the diffraction pattern disappears almost completely and the remaining intensity of a small facetted sample region dominates the pattern. Thus, Fe deposition reduces the reflectivity substantially and after ML completion the original Pt(997) periodicity of the topmost layer to which He scattering is exclusively sensitive is lost. This indicates a substantial roughening of the surface on an atomic scale without defined periodicity.

The sequence of measurements displayed in Fig.5 starts at Fig.5a with 1.5ML Fe deposited at 350K (same as Fig.4 bottom) and shows diffraction patterns recorded at 400K (Fig.5b) and 450K (Fig.5c) after the respective temperature has been established for more than 5 minutes. By annealing above 400K the narrow n=-3 diffraction peak of the ordered Pt(997) morphology re-appears. The result demonstrates that the defect-rich structure formed during Fe deposition at 350K orders at 400K due to the on-set of diffusion. This corresponds exactly to the temperature where a reflectivity maximum at 2ML coverage is observed (see Fig.3). The growth at low temperature can thus be identified with a Stranski-Krastanov (SK) like growth mode, which around 400K evolves towards a layer-by-layer growth mode. In the case of the vicinal Pt(997) surface this behavior is, however, induced by kinetics rather than by the balance of free surface and strain energies which are the important parameters in regular SK growth. A diffusion barrier for Fe which is much higher on the Fe covered surface than on the Pt substrate might, for example, explain the observation. It would lead to a reduction of step decoration in the second Fe layer in favor of island nucleation on the terraces.

Above 450K the He reflectivity curves in Fig.3 change their character completely. The maximum at monolayer coverage is replaced by a minimum. Above 600K, finally, the amplitude of the oscillations is strongly reduced and a continuous He intensity decrease becomes dominant. These changes are due to surface-restricted alloying between 450K and 550K and due to bulk alloying above 600K and will be discussed in section B.

The result presented so far emphasizes once more the role which steps play in the formation of adlayers. Fe growth on Pt(111) is reported to exhibit a three-dimensional growth mode with cluster formation at room temperature and 500K [6]. Pt(111) differs from the substrate used in this study essentially by its much lower step density and much larger average terrace size. The effect of an increased step density is to inhibit island nucleation on the terraces and favor the growth of smooth layers by means of preferential step decoration.

**B. Alloying of Fe with Pt from the Pt(997) substrate**

Fe and Pt exhibit a strong tendency for alloy formation and Fe exhibits antisegregation on Pt surfaces (segregation energy Fe/Pt: +0.63eV/atom; surface mixing energy +0.76 eV/atom [14]). Fe deposited on Pt(997) diffuses into the bulk on a timescale of minutes at temperatures T ≥ 600K. This is demonstrated in Fig.6 in which the Fe concentration is plotted as a function of annealing temperature. The concentration is obtained from Auger spectra ranging from 20 eV to 770 eV recorded for a primary electron energy of 3 keV. After deposition of 1ML Fe at T = 200 K, the temperature is increased in 50 K or 100 K steps to T = 800 K. AES measurements are carried out after each temperature step. The total dwell time at each temperature is 15 minutes, the next higher temperature is established more than 5 minutes before a measurement is started. The relative amount of Fe near the surface can be obtained by [15]

$$C_{Fe} = \frac{I_{Fe}}{S_{Fe}} \cdot \left( \sum_{\alpha=\text{Fe, Pt}} \frac{I_\alpha}{S_\alpha} \right)^{-1} \quad (1)$$

where $I_\alpha$ is the peak to peak intensity corresponding to a specific Auger line of the element $\alpha$ ($\alpha$ = Fe, Pt), and $S_\alpha$ is the relative sensitivity factor of the elements, namely $S_{Pt}$ = 0.025 and $S_{Fe}$ = 0.20 [15]. For the data displayed in Fig.6 the MNN peak at $E$ = 64 eV was used for Pt and the LMM peak at $E$ = 651 eV for Fe. The relative amount of Fe within the information depth of AES is constant within experimental accuracy for T ≤ 600 K, then starts to decrease and becomes undetectable above 850K. We assign the onset of the decrease at 600K to the beginning of Fe diffusion into the Pt bulk.

Bulk alloying thus explains the continuous decrease of the He intensity in Fig.3 above 600K. But it does not account for the shape of the He reflectivity curves at 500K and 550K. While the diffusion into the Pt bulk is still kinetically hindered below 600K there still is, however, the strong driving force for Fe to intermix with Pt [14]. This can drive an alloying process within the surface layer because the energy barriers at the surface are lower than for diffusion into the bulk. Alloying may e.g. proceed with a lowered activation barrier through Fe-Pt exchange at the step edges followed by diffusion *within* the terrace. We thus ascribe the initial slow decrease of He intensity in the reflectivity curves at 500K and 550K to the observation of alloying within the topmost layer. Clearly, this alloying process is not accompanied by He intensity oscillations. The oscillations observed at 500K and 550K set in only at a coverage above 0.5 ML and we find a minimum at a total Fe coverage of 1 ML and a maximum around 1.5 ML which corresponds to a shift by +0.5

ML with respect to the oscillations observed below 500K. This observation strongly suggests that the alloy formation is completed at a Fe coverage of 0.5 ML. When deposition continues beyond 0.5 ML a new Fe monolayer starts to grow on top of the alloy monolayer and results in the observed oscillations. An alloy formed by a Fe coverage of 0.5ML corresponds nominally to $Fe_{50}Pt_{50}$. In fact, the alloy with this stoichiometry is known to be reproducibly formed on Pt(111) and orders in alternating rows of Pt and Fe [16].

As a special case we want to discuss the Fe deposition at 500K in Fig.3 in more detail. At that temperature we first find a steep decrease of He reflectivity then a minimum close to 0.13 ML followed by an increase until 0.5 ML. Obviously, the slope at coverages below 0.13 ML corresponds to the low temperature behavior, for which no significant alloying occurs. At the completion of the first Fe row at the step edge (1 row = 1/8 ML) there is thus no indication of alloying yet. Only during continued deposition the data start to approach the 550K curve. Although the peak at 0.13ML in Fig.2 (which indicates step decoration) is primarily due to a specific morphology it is thus likely that the first row which decorates the step edge is formed mostly by Fe which only during continued deposition dissolves into the Pt terrace.

We now focus on the fact that for evaporation at T ≤ 400K an extremely low He intensity is observed at 0.5ML Fe coverage (Fig.3). This minimum cannot be ascribed to a highly disordered layer which re-orders when ML coverage is approached. It is well known from similar cases that destructive interference between atoms scattered from areas of different height plays an essential role [17]. The fact that this minimum disappears for T ≥ 550K corroborates our assignment to a transition from separated Pt and Fe regions to a homogenously mixed, i.e. alloyed, surface structure. We remark that for the surface of a bulk alloy, namely $Pt_{80}Fe_{20}$(111), substantial depletion of Fe in the top layers was found for temperatures above 700K [18, 19]. In contrast, Fe deposited on Pt(111) at room temperature followed by annealing between 600 K and 850 K is known to form an ordered surface alloy within the top-most layer containing almost 50% of Fe [5]. We suggest that the surface alloy which we observe for Fe/Pt(997) exhibits a structure similar to the (111) plane of the bulk $L1_0$ $Pt_{50}Fe_{50}$ alloy [16]. We could not obtain direct information on the chemical order of the surface alloy from the employed methods in the case of the vicinal surface. Neither He diffraction nor low energy electron diffraction (LEED) exhibit superstructure peaks corresponding to a doubled surface unit cell size. This can, however, be expected if there is only local chemical order which does not even extend to the nm range. In fact, already for the FePt alloy on Pt(111) which was annealed at much higher temperature no perfect nm-range ordering of the Fe rows was observed [16, 20].

As a next point we will discuss the alloying kinetics. The surface morphology obtained at temperatures below 500K which we attribute to the growth of a mostly pure Fe monolayer does not correspond to equilibrium morphology. We demonstrate this by interrupting the deposition at various coverages below 0.2ML, i.e. in the strongly decreasing part of the reflectivity curve (Fig.7). When we monitor the following increase of the He reflectivity signal as a function of time, we can study the alloying kinetics. We find a transition from long recovery times (>300s, i.e. much longer than deposition time of one atomic row) at 450K to short recovery times (≤100s i.e. shorter than deposition time for one row) at 500K. Within experimental precision the time constant is independent of Fe coverage. Fig.7 shows the decrease of He reflectivity for Fe deposition and the following He intensity recovery after deposition is stopped at (a) 0.06ML, (b) 0.13ML, and (c) 0.19ML at an intermediate surface temperature of 475K. By fitting single exponentials to the measured data (see Fig.7) a recovery time of 150s ± 20s is found for 475K. The recovery time constant changes substantially between 450K and 500K which suggests a large activation energy. Using an Arrhenius model we estimate an effective activation energy of 0.7eV±0.3eV. This barrier may reflect either the Pt-Fe exchange barrier or the diffusion barrier of Fe atoms (or the accompanying vacancies) inside the terraces.

Fig.7 contains also the continuous Fe deposition curves at 400K and 550K taken from Fig.3 for comparison with the deposition&recovery curves Fig.7 (a)-(c). During the initial Fe deposition the curves (a)-(c) closely follow the 400K curve. Upon stopping Fe deposition the relaxation towards equilibrium morphology becomes evident by the increase in He intensity. The data show that the recovery tends for long times approximately to He intensities on the 550K curve at the same Fe coverage which are marked as $a_\infty$, $b_\infty$, and $c_\infty$ in Fig.7. This observation strongly suggests that at 550K adsorption takes place close to equilibrium and the alloying process can be considered to be instantaneous on the time scale of Fe deposition. In contrast, the adsorption curves recorded between 400K and 550K are determined by the interplay of Fe deposition rate and alloying kinetics which in this range occur on the same time scale.

Finally, we want to compare the results presented in this paper to a study of Fe deposited on the low index Pt(111) by Jerdev et al. [5]. That study focuses on chemical surface composition for coverages between 1ML and 2ML analyzed by low energy ion scattering, X-ray electron spectroscopy, and LEED. Although both, the study by Jerdev et al. and ours, find a similar sequence of surface structures there are characteristic differences. Jerdev et al. [5] find surface alloying at a much higher temperature window (600K-850K). Temperature ramps are run on a much faster time scale (10s dwell time) compared to our experiments (typical dwell times of > 5 minutes). In addition, they deposit a given coverage and then anneal the sample while in the He intensity data Fe is deposited at the observation temperature. Qualitatively, these differences explain why the observed onset temperatures may be lower in our study. The major difference of the studies is, however, the high density of equidistant steps in our study which can activate a different path for alloying than on the Pt(111) surface. As already discussed, the uniform terrace width of 2nm provides a high density of high coordination sites for adsorbing atoms already at low adsorbate coverage. The residence time of single adatoms on the terraces is reduced with respect to Pt(111) and their attachment to substrate steps is enforced. The lowering of the observed alloying temperature thus suggests a more efficient alloying process for Fe at the Pt step edge with respect to Fe on terrace sites or at the edges of Fe islands.

## IV. CONCLUSION

We investigated the Fe growth on the vicinal Pt(997) surface in-situ by thermal energy He scattering. The results are summarized graphically in Fig.8. We find step decoration with separate Fe regions between 200K and 450K. A surface restricted alloy forms for deposition between 450 K and 600 K and bulk alloying occurs for T ≥ 600 K which is corroborated by Auger electron spectroscopy results. Comparison to the growth of Fe on Pt(111) demonstrates that a lower step density provides a rough film morphology whereas on the vicinal surface employed in this study a flat monolayer can be formed and three-dimensional growth can be suppressed at least up to monolayer coverage. Surface alloy formation appears on both surfaces but sets in at significantly lower temperature on the vicinal surface.

This study allows to derive three specific recipes to prepare nanometer-sized structures: 1) Perfect decoration of the step edges of Pt(997) by monatomic Fe chains occurs between 250K and 500K. 2) The best order of a Fe ML as judged from the He reflectivity is obtained around T = 400 K. 3) The FePt alloy remains restricted to the surface layer if Fe is evaporated in the range 500K to 550K. The surface alloy is assumed to consist of a few atoms long Fe chains embedded in the Pt layer.

**Acknowledgements**

One of us (T.Y.Lee) acknowledges financial support from the Korean Ministry of Education and Human Resources Development through the program of the Brain Korea 21 through Materials Education and Research division in Seoul National University.

**Figure Captions**

Fig.1 Sketch of the atomic arrangement of atoms on Pt(997) in side view (center) and top view (bottom). Definition of incident angle $\theta_i$ and total scattering angle $\chi$ in He atom scattering with respect to the macroscopic surface plane (top). The Bragg formula describes the position of diffraction peaks of order n for a given He wavelength $\lambda$ and surface periodicity d.

Fig.2 Normalized reflected He intensity as a function of Fe coverage monitored in-situ during deposition at the indicated substrate temperatures. He scattering in grazing geometry, wavelength $\lambda$= 1.01 Å, incidence angle $\theta_i$ = 85° and total scattering angle $\chi$ = 170°. For clarity the curves for T ≥ 200K are vertically shifted by +1 unit for each temperature step.

Fig.3 Normalized reflected He intensity as a function of Fe coverage in non-grazing geometry. He-beam incidence angle $\theta_i$ = 57.8°, $\chi$ = 101.8°. The presentation is equivalent to the one in Fig.2 except that for T ≥ 250K the curves are shifted vertically by multiples of 0.5 units.

Fig.4 He diffraction patterns in non-grazing geometry after Fe deposition at T = 350 K. The Fe coverage is indicated at each trace. He beam incidence angle of $\theta_i$ = 57.8°. Peak positions are indicated as dashed lines and correspond to the diffraction orders n = -4 to n = -2. Please mark the logarithmic intensity scale. For clarity, the upper curves are multiplied by the indicated factors.

Fig.5 He diffraction patterns in non-grazing scattering geometry after deposition of 1.5ML of Fe at T=350K (a). The following measurements were made successively after heating the sample to (b) 400K and (c) 450K. For comparison, (d) is the diffraction scan of a well prepared clean Pt(997) surface. The scattering geometry is the same as for the measurements in Fig.4. Please mark the logarithmic intensity scale.

Fig.6 Relative amount of Fe monitored by Auger spectroscopy while ramping the sample temperature in steps of 50K or 100K. Initially, 1ML of Fe is deposited at 200K. The time between two measuring points is 15 minutes. The relative amount of Fe is obtained from eq. (1) using peak intensities from Auger spectra (see text). The dashed line is a guide to the eye.

Fig.7 Deposition of Fe (coverage scale, above) and recovery of reflected He intensity (time scale, below). In the curves (a)-(c) Fe is deposited at 475K at a deposition rate of 0.0014 ML/s. When deposition is stopped at the coverages (a) 0.06 ML, (b) 0.13 ML, and (c) 0.19ML the curves switch from a negative to a positive slope, indicating recovery of He reflectivity. The solid lines in the curves (a) to (c) are exponential fits. For comparison we show the continuous deposition curves at 400K and at 550K taken from Fig.3. At 400K no alloying occurs and at 550K alloy formation occurs in equilibrium with deposition as discussed in the text.

Fig.8 Diagram representing the growth mode of Fe on Pt(997) as a function of temperature at a deposition rate around $10^{-3}$ ML/s.

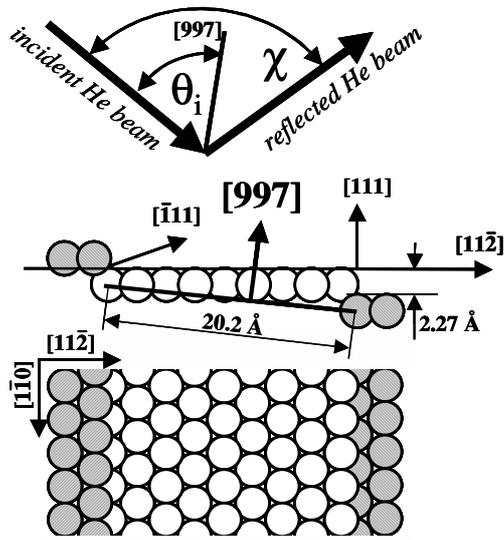

**Fig. 1 Lee *et al*.**

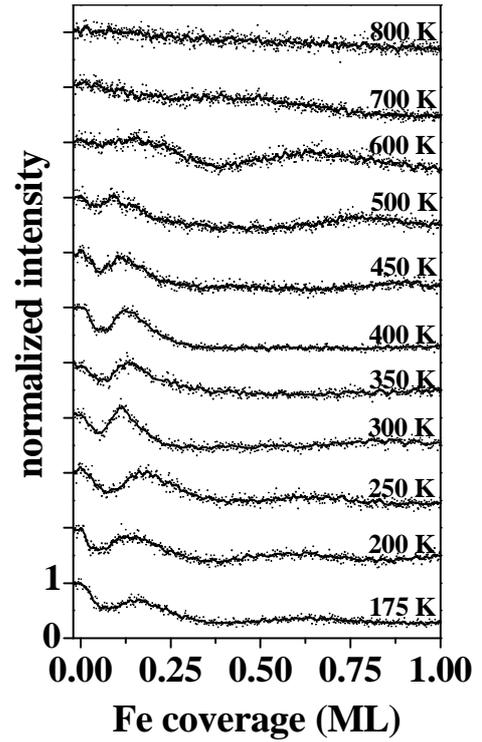

**Fig. 2 Lee *et al*.**

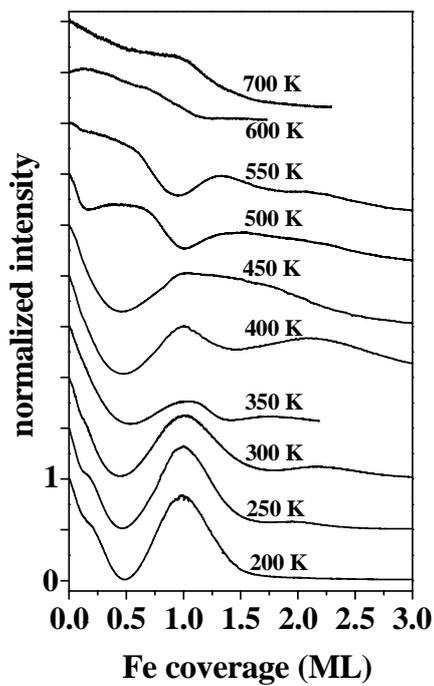

**Fig. 3 Lee *et al*.**

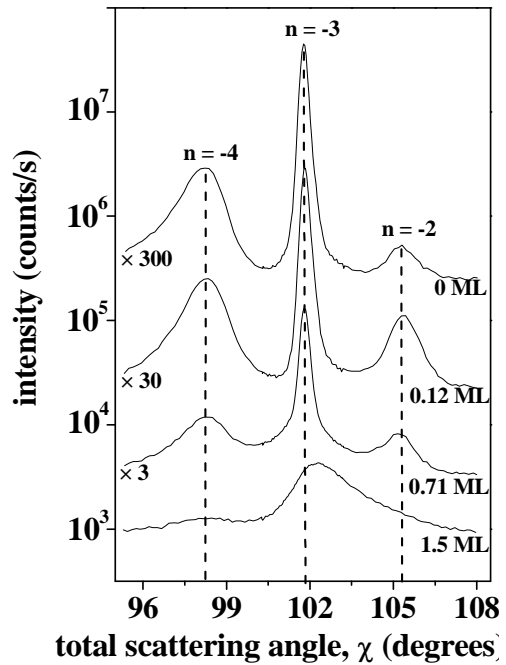

**Fig. 4 Lee *et al*.**

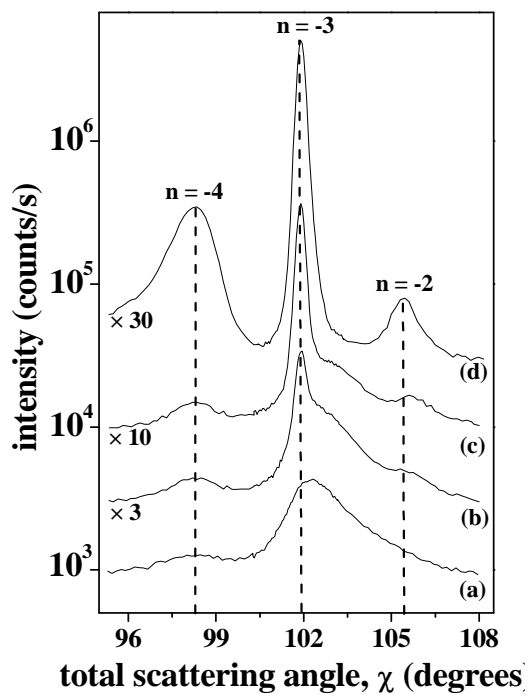

**Fig. 5** Lee *et al*.

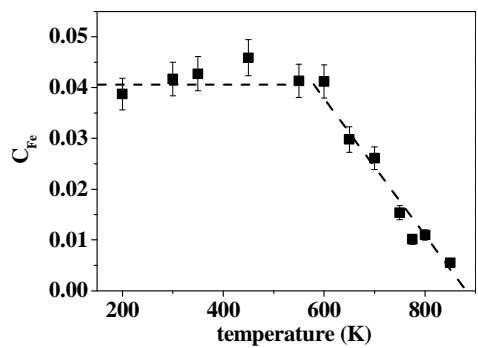

**Fig. 6** Lee *et al*.

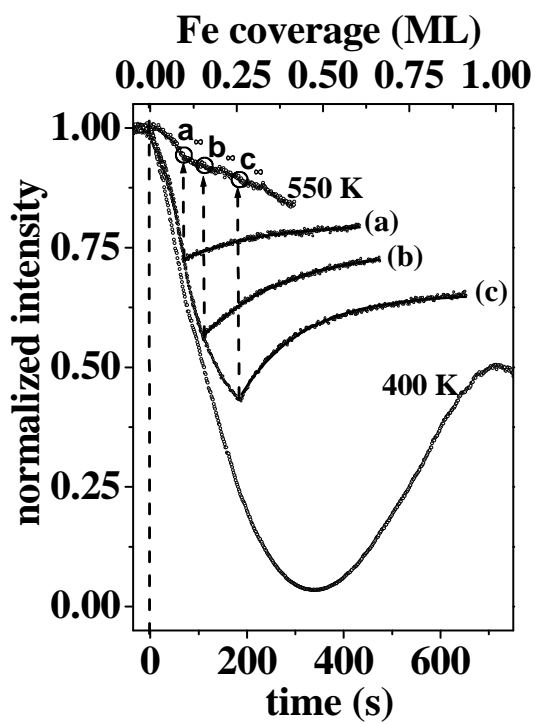

**Fig. 7** Lee *et al*.

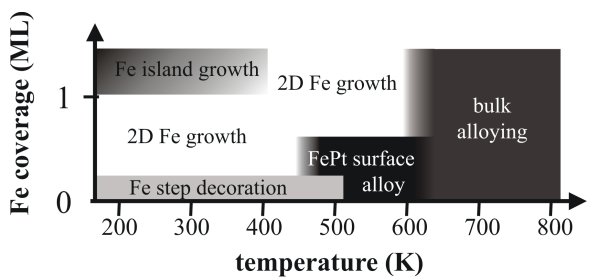

**Fig. 8** Lee *et al*.

14